# Using Experimental Vignettes to Study Early-Stage Automation Adoption

**Sarah Janböcke**
University of Siegen
Siegen, 57072, Germany
sarah.janboecke@uni-siegen.de

**Diana Löffler**
University of Siegen
Siegen, 57072, Germany
diana.loeffler@uni-siegen.de

**Marc Hassenzahl**
University of Siegen
Siegen, 57072, Germany
marc.hassenzahl@uni-siegen.de

## Abstract
When discussing the future of work and in detail the concerns of workers within and beyond established workplace settings, technology-wise we act on rather new ground. Especially preserving a meaningful work environment gains new importance when introducing disruptive technologies. We sometimes do not even have the technology which effects we are willing to discuss. To measure implications for employees and thus create meaningful design variants we need to test systems and their effects before developing them. Confronted with the same problem we used the experimental vignette method to study effects of AI use in work contexts. During the workshop we will report our experiences.

## CSS Concepts
• Human-centered computing~Human computer interaction (HCI)   • Human-centered computing~User studies  • Human-centered computing~Field studies

## Introduction
Many employees fear that a substantial amount of work will be automated away and that remaining jobs substantially change in the near future [3,4,6,11–13]. These concerns are major forces in discouraging adoption of labor-saving technologies such as artificial intelligence (AI)-enabled autonomous systems. In contrast, automation holds opportunities by carrying out mundane and repetitive work, leaving workers with supposedly more valuable duties [10]. But what exactly is valuable and meaningful to those having to work with an autonomous system?

So far, automation design primarily focused on optimizing operational efficiency and safety by reducing human involvement as well as making systems easier to use for the operator [10]. These approaches often completely change the nature and structure of tasks, resulting in low acceptance, more severe failures and an increased erosion of the sense of purpose given by work [10]. More recent notions of Experience Design, Positive Design and Design for Wellbeing therefore argue that technology should not only be made usable but crafted in a way that actively contributes to meaningful, fulfilling work [5,7,8]. Placing workers' wellbeing at the center of design efforts means that AI-enabled autonomous systems have to be created to support meaningful work practices. This requires the exploration of design variants of a technology which does to some extent not yet exist. In an effort to identify potential key sources of meaningful work with an AI-enabled autonomous system with reasonable efforts, we used the method of experimental vignettes. We describe the method, findings and lessons learned regarding its suitability to study early-stage automation adoption in the following.

## Experimental Vignette Methodology

Vignette studies present comparable short descriptions of situations or persons, the vignettes, to participants within surveys in order to get feedback on their judgement, feelings or attitudes about these scenarios [9]. Atzmüller and Steiner describe vignettes as a "short, carefully constructed description of a person, object, or situation, representing as systematic combination of characteristics." [2, p.128]. These characteristics vary throughout different vignettes and represent the levels of theoretically important constructs or independent variables in the study. Participants in vignette studies are normally confronted with either single vignette or a set of different vignettes depending on the study design [11,13,12]. The methodology originated in sociology where vignette studies were used frequently to balance the high validity of experimental research with the high external validity of non-experimental research like survey methods [12,13] Vignette studies prove to be rather cost-effective and are used when rapid data collection is needed. Moreover, as the scenarios presented in vignette studies do describe a certain situation in detail, participants do not necessarily need in-depth knowledge, although this increases the validity of the results.

## A Case on Early-Stage Automation Adoption

We carried out a series of three online vignette studies on the effects of AI use on employees' work satisfaction. For each study between 60-120 participants were recruited via business-relevant networking sites. The vignettes described a specific work context and how the interaction with a future AI-based system could be like, for example:

My boss commissions me by mail with the solution of an important problem. I assign a centrally available, intelligent software with the further analysis and elaboration. This software authors a written solution concept and sends it to me. I check it and send it to my boss. My boss rates the proposed solution as effective and compliments me for the preparation.

The scenarios were rated in terms of work satisfaction on multiple Likert scales. In our first study we showed the linear connection between increasing automation and decreasing employees' satisfaction. Thus, we started two further in-depth studies in which we focused on different levels of automation and its effects on meaningful work environments and artefacts.

We chose this approach to access participants' attitudes that are difficult to anticipate in a classical questionnaire study when testing future scenarios that do not recline to the participants' work realities. Testing with this method turned out to be a cost- and time-efficient way to reach a high number of participants, especially since the target group was difficult to recruit due to work accountabilities, i.e., employees with a certain degree of experience in working with digital technologies. Furthermore, the vignette study gave us the opportunity to explore technology scenarios and contexts that would have been hard to test when not properly developed yet or even non-existent.

Since the method is comparable lightweight, we assumed to find a high willingness to participate. The vignette method does not require any preparation from the participants. Even during the questioning, the participants do not need to explicitly deal with the problem but just „feel" themselves into situations [9]. So, we assumed the hurdles for participants to be small and to reach a rather small dropout rate. Our participants could easily figure out the situations described in the vignettes. In our first survey we added a comprehension question in which we asked if the described scenario was understandable and easy to anticipate. We found that the majority easily anticipated and imagined the outlined future AI-enabled system support, which in fact is currently relatively difficult to grasp.

One limitation of the method is the reduced complexity of the real world within the vignette descriptions. The vignette design therefore only allows for a superficial view on work-relevant details. The same applies to the low complexity of the answers where it was not possible to lighten detailed facets as in work relevant questionnaires or qualitative interviews typical in the field of work psychology and sociology. Thus, we could not gather work-relevant constructs in-depth but instead an overall tendency, e.g., which system design variants are likely preferred and contribute to wellbeing, which is nevertheless valuable in early stages of the technology development process.

Furthermore, even if we deal with an imagination method in which a future scenario is to be anticipated we still have to assume that it is more a cognitive than an experience-based method. Participants do not experience the scenario in a tangible way, but read it on a tablet or computer. Thus, it is a concept-based, but still primarily cognitive method.

Finally, we felt that we had to be very careful with the wording of the vignettes, so that the situations could be clearly understood without too much room for interpretation.

## Conclusion

Our conclusion from the experiences with the vignette methodology in a series of case studies on automation design is that the method provided valuable insights to the system design at a very early, i.e., pre-development, stage. From this perspective, we can recommend the method to explore crucial system parameters and their effects on workers wellbeing before resources are spent in a costly development process with missing effectiveness or even negative effects as result. We still recommend a very intensive examination of the pros and cons of the method, which comes primarily from sociology and was not originally intended for research in human-computer interaction, in order to be able to rule out possible hurdles or potential validity threats.

**Short Author Bios**

Sarah Janböcke is scientific researcher at the Ubiquitous Design group at University of Siegen and works as digital leadership consultant for various in industry partners. She has a background in Industrial Design and Work and Organizational Psychology. Her main interest is to create humanized work environments especially in times of fast developing technology integration in work contexts.

Diana Löffler is postdoc at the Ubiquitous Design group at University of Siegen. She has a background in psychology and human-computer interaction and is interested in the design of autonomous systems with a focus on artificial intelligence and robotics.

Marc Hassenzahl is Professor for Ubiquitous Design at University of Siegen. He is interested in designing meaningful moments through interactive technologies – in short: Experience Design.

**Acknowledgements**

This work was partially supported by the German Federal Ministry of Education and Research (BMBF), project GINA (Grant: 16SV8095).